# publications

*Review*

# Research misconduct – definitions, manifestations and extent

**Lutz Bornmann** [1]

[1] Division for Science and Innovation Studies, Administrative Headquarters of the Max Planck Society, Hofgartenstr. 8, 80539 Munich, Germany; E-mail: bornmann@gv.mpg.de; Tel.: +49-89-2108-1265

**Abstract:** In recent years, the international scientific community has been rocked by a number of serious cases of research misconduct. In one of these, Woo Suk Hwang, a Korean stem cell researcher published two articles on research with ground-breaking results in Science in 2004 and 2005. Both articles were later revealed to be fakes. This paper provides an overview of what research misconduct is generally understood to be, its manifestations and the extent to which they are thought to exist.

**Keywords:** misconduct, plagiarism, fabrication and falsification of data

## 1. Introduction

A specific achievement expected of the natural and technological sciences and the social and financial sciences is that they make accurate statements about the world in which we live [1]. To offer this achievement reliably to society, the formal processes used to communicate these insights must be designed to be not only cumulative and complementary but also trustworthy and reliable [2,3]. The publication of a paper in a journal is an invitation to other scientists to use the information it contains (the ideas and findings) for their own research. With the quality of their own work in mind, these scientists must be able to be certain that their colleagues' research is solely directed at making correct statements about natural or social phenomena [4]. Luhmann [5] calls this 'system trust' which allows scientists to avoid the immense effort of checking the reliability of the results of their colleagues' research before they use them [6].

Given the serious cases of research misconduct which have been uncovered in recent years, some people (scientists, editors and politicians) have expressed doubt that science can still operate today in an atmosphere of 'system trust' [7]. For example, Woo Suk Hwang, a Korean stem cell researcher published two articles on research with ground-breaking results in *Science* in 2004 and 2005 [8]. "Both [papers] have turned out to be complete and deliberate fakes" [9, p. 122]. Four more cases of research misconduct which attracted weeks of attention in the general media (particularly in Germany) in the last few years are those of the cancer researchers Friedhelm Herrmann and Marion Brach, the physicist Jan Hendrik Schön [10], the anesthesiologist Joachim Boldt [11,12] and the psychologist Diederik Stapel [13]. Research results were massaged, images in scientific papers faked and research proposals from colleagues recommended for rejection and subsequently submitted as the wrongdoer's own. These and similar cases wasted the time and research funds by other scientists until the fraud was ultimately detected and a retraction published. Taking these and other cases as examples, the public and the scientific community discussed extensively whether they were rare isolated cases or typical researcher behaviour. The greater the extent of the fraud and deception, the less likely it seemed that science would be able to operate in an atmosphere of trust.

Research misconduct has a long history in science [10]. In the USA, the subject has been in the public domain since numerous cases of fraud were uncovered at respected US research institutions during the 1970s and 1980s [14]. According to Garfield's [15] historiography, which is published at the URL http://garfield.library.upenn.edu/histcomp/index-plagiarism.html (accessed: 12.08.2013), around 1000 publications in total have discussed the subject of 'Misconduct in Science' since the 1970s. They were mainly published in English-speaking countries and were mostly editorial material and letters; research articles are rare. In addition to these publications, there are many monographs, collected writings and grey literature of which Garfield's historiography [15] takes no account. All in all there are very few empirical studies which have applied systematic analysis to the issue of 'Misconduct in Science' [16]. This is astonishing given the resonance which regularly greets the subject – both within and outside of the science community.

## 2. Definition of research misconduct

There is no standard, generally valid definition of research misconduct. Gilbert and Denison [17] provide an overview of a number of definitions of research misconduct as formulated by various international scientific institutions. For example, the National Science Foundation (NSF, Arlington, VA, USA) offered a very general definition in the 1990s. It considered misconduct to be a serious deviation from generally accepted practices in a certain discipline [18]. The advantage of such a vaguely formulated definition is that it allows more freedom with which to proceed against aberrant conduct. If what constitutes misconduct is formulated very specifically, it can later transpire that a new form of deception cannot initially be classified as misconduct nor can sanctions be imposed as appropriate. However, in 2000 the NSF dropped the vaguely formulated definition when the Office of Science and Technology Policy (OSTP, Washington, DC) developed a common federal definition of misconduct.

This definition, which many other scientific institutions use and which is given as the basic definition in many publications on research misconduct, is that of the OSTP: "*Research misconduct is defined as fabrication, falsification, or plagiarism in proposing, performing, or reviewing research, or in reporting research results*" [19]. The definition names specific conduct with which a researcher commits an offence against the rules of good scientific practice. For German-speaking countries, the German Rectors' Conference (HRK, Cologne) has provided the following definition as part of its recommendations on how to deal with research misconduct: Research misconduct is the provision of incorrect information in a scientific context of substantial significance, either intentionally or grossly negligently, the infringement of intellectual property of others or harm done to their research in any other way [20].

Recently, Fanelli [21] published a commentary on research misconduct in *Nature*, arguing that misconduct should be redefined as "biased reporting," of all sorts: Misconduct should be defined as "any omission or misrepresentation of the information necessary and sufficient to evaluate the validity and significance of research, at the level appropriate to the context in which the research is communicated." This definition is closely connected to the philosophy of science based on Popper [22]. According to Fanelli [21] "scientific knowledge is reliable not because scientists are more clever, objective or honest than other people, but because their claims are exposed to criticism and replication."

The literature on fraud and deception refers frequently to some factors which are considered important for classifying certain behaviours as misconduct:

- *Non-norm-compliant conduct*: According to Mayntz [23] research misconduct is when one of the norms of the research system is violated intentionally or negligently.
- *Intentional misconduct*: Science is undertaken in a culture in which the intention (even legally) makes a big difference to the moral assessment [1]. Accordingly, it is frequently stressed that it must be possible to attribute intention to misconduct; generally speaking, it is *not* a case of unintentional error or carelessness during research or the interpretation of the findings [24]. Scientific research is always prone to error (particularly where research is innovative or cutting-edge) [1]. Science is even closely associated with the right to use fallible trial and error, to take risks and unrecognised (unorthodox or intuitive) routes in research, even to have imbalance and incompleteness: simply to attempt to find out the truth [25].

  However, if a scientist's work is flawed and his conduct can be considered grossly negligent, even unintentional faults are deemed to be misconduct [16]. Misconduct can even arise when the researcher is completely convinced that his distorted or selective interpretation is correct [23].
- *Research with significant consequences*: For Fuchs and Westervelt [16] a necessary requirement for classifying conduct as deviant is that it is associated with a significant effect on the research and its results. Minor transgressions which do not have an impact on the process of building scientific knowledge should therefore not be classified as misconduct.
- *Academic research*: Taylor [26] points out that a number of behaviours which are considered misconduct in academic research are common practice in industrial research (such as using other people's unpatented ideas for a company's own commercial purposes).

As the boundaries between unintentional faults, carelessness with serious consequences, gross negligence and intentional deceit are unclear, in many cases it is a difficult undertaking to classify certain behaviours as misconduct [27].

## 3. The manifestations of research misconduct

There are many different manifestations of research misconduct [23] which occur at every stage of the research process (data generation, recording, review and publication / dissemination of scientific knowledge) [28]. The typical manifestations of misconduct are described below (those which are discussed most frequently in the literature). Very comprehensive overviews of the various manifestations of misconduct have been published by the Parliamentary Office of Science and Technology [29] and by Helton-Fauth, *et al.* [30]. The overview provided by Helton-Fauth, *et al.* [30] is particularly interesting because the authors analysed the content of the standards for good scientific practice published by various scientific institutions in recent years (or the markers for the violation of these standards).

The fabrication and falsification of data are behaviours which are most frequently associated with research misconduct [17]. Fabrication means that data is quite simply invented. Falsification means that existing data is 'pruned' to take on the required form or 'massaged' to give the desired result [31]. 'Pruning and massaging' can be undertaken through the use of inappropriate methods of data analysis [29], the (tacit) exclusion of outliers in data analysis [17] or the unpermitted manipulation of graphics (with software, such as Photoshop) [32]. As the manipulation of graphics is a relatively frequent problem in journal manuscripts, some editors already employ specialists to examine graphics for unpermitted (or still permitted) manipulation [27,33].

The second major area typically associated with research misconduct is plagiarism: passing off someone else's intellectual property (information or ideas) as one's own achievement without giving the actual source (e.g. in research papers) [27]. Particularly in the era of the Internet, this form of misconduct is acquiring great significance: "There is now an enormous amount of information available via the Internet; text is very easy to copy and paste, and ideas can be gleaned from a multitude of sources" [27, p. 177]. While some authors consider 10 copied words in a text to be plagiarism, others require at least 30 [17]. Unlike inventing and falsifying data, plagiarism does not undermine the credibility of scientific statements [23]. However, since claims and entitlement to priority play a significant role in the bestowing of accolades in science [34], plagiarism presents a problem for the reward system it operates. Inventions and discoveries are a scientist's most valuable capital [1]. "We do not make products, we make ideas. Steal my words, and you steal my authorship. Steal my idea, and you steal my identity as a scientist" [14, p. 212]. Yet there is evidence that word plagiarism can be rather innocent. According to Cameron, *et al.* [35] the fact that "the majority of the scientists publishing in English-language journals are not native English speakers … has important implications for training concerning ethics and enforcement of publication standards, particularly with respect to plagiarism" (p. 51).

As well as plagiarism and the invention or falsification of data as the typical fraudulent and deceptive behaviours, the following forms of misconduct are also mentioned quite frequently in the literature: redundant, multiple or broadly overlapping publishing [27], the publication of research results in very small units – known as 'salami slicing' [17], dual or multiple submission of manuscripts ('shot-gunning'), listing authors on a publication who have not made a substantial contribution ('gift authorship'), not listing authors who have made a substantial contribution ('ghost authorship') and listing co-authors against their will [36]. Generally, a difference is made in these various forms of misconduct between minor misdemeanours and serious violations. For example, the German Rectors' Conference [20] in Germany classifies the following five manifestations as serious misdemeanours: 1) Falsifying information, 2) infringement of intellectual property, 3) claiming another person as a (co-)author without their permission, 4) sabotaging research work and 5) destroying primary data.

Giles [37] notes concerning the seriousness of misdemeanours that the mere invention of data and results, as practised by the Korean stem cell researcher Woo Suk Hwang (see above) was morally more repugnant than slightly manipulating an illustration, but minor acts of misconduct are much more common, and potentially more damaging to scientific progress. This assessment is supported by the results of two studies. A Delphi survey of scientists on misconduct in clinical trials found that: According to this expert group, the most important forms of research misconduct in clinical trials are selective reporting and the opportunistic use of the play of chance. Data fabrication and falsification were not rated highly because it was considered that these were unlikely to occur [38, p. 331]. de Vries, *et al.* [39] had asked scientists which form of misconduct in their view impacted most strongly on the search for and creation of reliable knowledge. "We found that while researchers were aware of the problems of FFP [falsification, fabrication, and plagiarism], in their eyes misconduct generally is associated with more mundane, everyday problems in the work environment. These more common problems fall into four categories: the meaning of data, the rules of science, life with colleagues, and the pressures of production in science" (p. 43).

## 4. The extent of research misconduct

We can expect to find answers to the question of the extent of research misconduct offered by those institutions who are asked for guidance in actual cases of fraud and deceit [40]. In Germany, the German Research Foundation (DFG, Bonn) created the post of ombudsman in 1999 in the wake of a spectacular case of fraud. The ombudsman's function is to be an institution for advice and mediation, which means that he/she does not consider himself/herself an investigating body for uncovering research misconduct, but as an advisor to scientists and a mediator in any conflicts that arise [41]. There are similar institutions in other countries which can also investigate and impose sanctions, in addition to having an advisory role (such as the Danish Committees on Scientific Dishonesty, DCSD, Copenhagen).

In 2012, 59 cases of research misconduct were referred to the DFG's ombudsman (mostly accusations of misconduct concerning authorship), of which 19 received attention [42]. The ombudsman did not take action in many cases, because the cases were already addressed elsewhere. In the UK, the Committee on Publication Ethics (COPE, London) dealt with 212 cases from 1997 to 2004; 163 of them were very probably instances of misconduct [43]. The most common manifestations of misconduct were: duplicate/redundant/salami publication (n=58), authorship issues before or after publication (n=26), and no ethics approval (n=25). The Office of Research Integrity (ORI, Rockville, MD) in the US which investigates cases in research projects financed by the National Institutes of Health (NIH, Bethesda, MD, USA) was brought 267 accusations of misconduct in 2006. Of the 267 cases "71 were assessed by ORI in detail for a potential inquiry or investigation; 29 assessments resulted in the opening of formal cases. Of these, 23 were from 2006 allegations and 6 were from previous year allegations. One 2005 case developed into two cases in 2006 when a second respondent was identified. In total, 20 allegations were administratively closed" [44, p. 3]. A detailed examination of ORI reports from 1992 to 2003 can be found in Kornfeld [45]. In 2005, the DCSD handled 7 cases of which none were classified as research misconduct.

Using similar figures from ago, Fuchs and Westervelt [16] tried to estimate the actual extent of research misconduct in the formal science communication processes. Their calculations indicated that only around 0.01% of publications are associated with it. Other similarly low figures have been presented by other authors [46]. These estimates seem to confirm Merton's [31] assessment that fraud and deceit in science are extremely rare and that they are unusual rather than typical. Merton [31] attributes this to the effect of scientific norms in the research process, the unique combination of competition and trust between scientists and their mutual dependence in the production of their research results [47]. Furthermore fraud and deceit – according to Popper's philosophy of science – do not have a sustainable chance of survival in the evolutionary contest represented by scientific findings [48]. A study is forgotten when its results are not replicated and it loses its claim to truth.

A number of scientists have however expressed doubt in recent years that research misconduct only occurs in isolated cases which play a merely marginal role in the process of building scientific knowledge. In their view, the uncovered cases represent just the visible tip of a huge iceberg which is barely detectable below the surface. They give these reasons for this claim: (1) When scientists encounter research results which they do not believe are plausible or reproducible, they tend to suspect a mistake in the study and not misconduct [6]. (2) If mistakes are found in a study, they are usually not recorded or documented for public access [49]. (3) On the one hand the social proximity of scientists in a research group is frequently an important condition for the discovery of misconduct; on the other, it is also a typical reason to prevent it being anticipated and reported [50]. (4) The blurred boundaries between innocent error, avoidable faults, intentional 'bending' and massive falsification [1] often make it difficult to determine misconduct. According to the findings of Garcia-Berthou and Alcaraz [51], errors in scientific publications are very frequent and it is not clear whether they are the result of carelessness or deliberate falsification [52].

In order to gain an impression of the actual extent of research misconduct, some studies have been carried out during the past few years which have looked at the estimates of (mis)conduct among employees or colleagues by the person being surveyed. The studies are generally based on non-representative samples. A survey of 1645 coordinators of clinical trials in the USA showed that a fifth of those asked had already encountered research misconduct at work [53]. Swazey, *et al.* [54] asked approximately 4000 scientists about their experience of research misconduct among their colleagues. Slightly fewer than 10% of those surveyed said that they knew of at least one case where data had been plagiarised or falsified. In a survey of 300 cancer researchers whose application to the NIH had been rejected, around 20% claimed that the information in their proposals had been misappropriated [55]. Wilson, *et al.* [56] interviewed Research Integrity Officers at 90 major research universities in the US. Of a total of 553 checks that the RIOs had carried out at their institutions, around 38% revealed problems with the documentation of research. When we compare these figures from four studies in which the conduct of colleagues and employees is appraised with those available from institutions who provide advice and help in the event of misconduct (see above) we can assume that the cases dealt with by the institutions are registered isolated cases of a phenomenon which is more widely distributed in research.

The results of the most elaborate study so far on the extent of research misconduct were published in 2006 in *Nature*. "We are the first to provide empirical evidence based on self reports from large and representative samples of US scientists that document the occurrence of a broad range of misbehaviours" [57, p. 737]. The authors wrote to approximately 8000 scientists who were at the start of or halfway through their careers. The response rate was around 50%. Roughly one third of those surveyed (33%) said that they had exhibited at least one instance of misconduct in one of its more serious manifestations (such as "falsifying or 'cooking' research data, using another's ideas without obtaining permission or giving due credit, failing to present data that contradict one's own previous research", p. 737). Martinson, *et al.* [57] interpret their results thus: "Our findings suggest that US scientists engage in a range of behaviours extending far beyond FFP [falsification, fabrication, and plagiarism] that can damage the integrity of science" (p. 738).

Fanelli [58] published the first meta-analysis of surveys on research misconduct. The term meta-analysis was coined by Glass [59] to refer to "the statistical analysis of a large collection of analysis results from individual studies for the purpose of integrating the findings" (p. 3). The meta-analysis of Fanelli [58] was limited to misconduct that distort scientific knowledge (e.g., fabrication, falsification, and "cooking" of data). That means among other things that results on plagiarism were excluded. Eighteen studies have been included in the meta-analysis. The most important results of the meta-analysis are as follows: "A pooled weighted average of 1.97% (N = 7, 95% CI: 0.86 - 4.45) of scientists admitted to have fabricated, falsified or modified data or results at least once – a serious form of misconduct by any standard – and up to 33.7% admitted other questionable research practices. In surveys asking about the behaviour of colleagues, admission rates were 14.12% (N = 12, 95% CI: 9.91 - 19.72) for falsification, and up to 72% for other questionable research practices."

Even though most of the studies on misconduct by scientists were carried out very carefully, given the subject matter, the question of whether the results reflect the genuine extent of research misconduct remains. For example, we can assume with some certainty that despite being assured anonymity, not all those involved would have admitted to having falsified data [48]. We can also assume that those for whom fraud and deceit are part of everyday research life were precisely those who did not take part in the survey. Some authors recommend the Randomised Response Technique (RR Technique) as a solution to not only these problems, but to those generally associated with any (self-)survey of misconduct. It can assure those surveyed of complete anonymity [48,60,61]. However, in a survey of economists on research misconduct, using this method did not deliver the hoped-for advantage compared to a direct question [62]. Approximately 4.5% of those surveyed with the RR technique and of those asked directly, i.e. without any additional anonymisation, said that they had already falsified data. The response rate in both groups was about the same, at around a quarter: "a surprising result in light of the additional protection that the RR method offers" [62, p. 165]. Another problem in surveys which cannot be addressed by the RR Technique is its inability to overcome self-delusion about one's behavior. Those scientists who habitually misbehave might have been less likely to respond to a survey, but among those who did, they may also be differentially unlikely to perceive themselves that their behaviors were wrong.

Another method which can be helpful to reveal possible biases affecting scientific research is meta-analysis (see above). Fanelli and Ioannidis [63] extracted more than thousand primary outcomes appearing in nearly 100 meta-analyses published in health-related biological and behavioural research. They measured how individual results deviated from the overall summary effect size within their respective meta-analysis. They found that "primary studies whose outcome included behavioral parameters were generally more likely to report extreme effects, and those with a corresponding author based in the US were more likely to deviate in the direction predicted by their experimental hypotheses, particularly when their outcome did not include additional biological parameters. Non-behavioral studies showed no such 'US effect' and were subject mainly to sampling variance and small-study effects, which were stronger for non-US countries. "Results like those of Fanelli and Ioannidis [63] point to problems in scientific research where, for example, career systems are too heavily oriented toward productivity, such as those adopted in the US. The results are moving us away from a "regulatory compliance" and criminological way of viewing how to think about and address misconduct, and more towards policy that takes better account of human frailty (e.g., cognitive biases and influences of the environment on our behavior). It moves us towards a policy stance that is more focused on treating these problems as issues of quality management in science.

## 5. Conclusions

Generally, the public outside of the scientific community associates research with a high level of moral integrity. "Scientists are generally perceived as well-intentioned seekers of truth; universities, as cathedrals of learning and as producers of knowledge vital to the health and welfare of society" [14, p. 211]. Against the background of such idealised images, every spectacular case of fraud which is discovered and discussed at length in the public media seriously damages the trust placed in science [17,27] and casts doubt on the entitlement of researchers to financial support [23]. Regarding science itself, one can assume that the spectacular cases of fraud have hardly any negative effect on the process of building knowledge with research [1]. They seem to occur too seldom and the lack of replication seems to put a question mark over the "massaged" results. With regard to the frequently postulated loss of trust in science caused by spectacular cases of fraud, one must distinguish between its internal processes and its external image.

More minor transgressions, where the rules of good research practice are knowingly or unknowingly (grossly) disregarded in the everyday routine in the laboratory, are instead seen to present a risk to the process of knowledge-building. If one is more concerned about the performance of science as an enterprise rather than about a moral judgement, then a critical eye should be cast less at the relatively few cases of obvious falsification and much more at partial falsification and sloppy practice in research [1]. However, because the figures published on the frequency with which the individual manifestations of research misconduct occur vary greatly (see above) and the reliability of these figures (particularly where researchers are asked directly) is in very great doubt, the extent of the risk posed by the smaller misdemeanours to the progress of scientific knowledge is almost impossible to estimate. There is as yet no suitable method with which to obtain reliable and meaningful figures.

## Acknowledgements

The author declares no conflict of interest. I am grateful to the external editor and three anonymous reviewers for their very helpful comments to improve the manuscript.